\documentstyle[12pt,epsfig]{article}
\def\beq{\begin{equation}}
\def\eeq{\end{equation}}
\def\beqa{\begin{eqnarray}}
\def\eeqa{\end{eqnarray}}

\begin{document}

\begin{flushright}
ITP-SB-97-43\\
\end{flushright}

\vspace{3mm}
\begin{center}
{\Large \bf SUMMARY, DIS 97\footnote{Presented
at the Fifth International Workshop on Deep Inelastic
Scattering and QCD, Chicago, IL, April 14-18, 1997.}} 
\end{center}
\vspace{2mm}
\begin{center}
{\large George Sterman}\\
\vspace{2mm}
{\it Institute for Theoretical Physics\\
State University of New York at Stony Brook\\
Stony Brook, NY 11794-3840, USA} \\  
\end{center}

\begin{abstract}
Some of the experimental and theoretical results discussed at
the Fifth International Workshop on Deep Inelastic 
Scattering and QCD are reviewed.
\end{abstract}

\section{Introduction}

Each ``DIS" workshop in this series has reflected 
exciting developments
from the foregoing year in both theory and experiment.
The Organizing Committee and Argonne National Laboratory
have made possible another such meeting, 
distinguished by reports of progress in
many directions, reflecting a vibrant and expanding field.
DIS '97 was favored as well by strong participation
from the hadron and ${\rm e}^+{\rm e}^-$collider communities.  
The timing for this participation could not have
been better, given the fruitful interplay between these complementary
arenas in the investigation of the strong interactions.  We shall see
examples repeatedly in what follows.

Let me begin with a summary of the Summary -- a set  of impressions
of the past year for deeply inelastic scattering (DIS),
as represented at this workshop.  
1996-97 saw the culmination of a set of classic 
experiments, including final or near-final analyses from
NMC \cite{nmc}, CCFR \cite{ccfr}, SMC \cite{smc}  and E154 \cite{E154}.  
Each has illuminated hadron structure \cite{Robplen}
in a memorable fashion.  From HERA \cite{ZEUSplen,H1plen}, 
we heard of increased
coverage in the kinematic range in $x$ and $Q^2$, including reports on 
an excess of ``large-$x$ and $Q^2$" events 
\cite{H1highQ,ZEUShiQ}, which has drawn the attention of the press.
The newly-rich study of DIS jet cross sections also affords 
a wealth of predictions that can be cross-checked in
hadron-hadron collisions \cite{Montplen}.
At the same time, a ``coming-of-age" of
diffraction and photoproduction \cite{Forplen} is making possible
for the first time truly
quantitative studies of these processes, at the boundary of
perturbative and nonperturbative dynamics \cite{Dokplen}.  
Finally, early results from HERMES are appearing in the fast-developing 
program of spin physics in DIS and elsewhere \cite{Milplen}.

Theoretical developments \cite{Robplen,Dokplen} 
reported at the conference were characterized by 
the requirements of  ``precision
QCD", a concept almost unimaginable fifteen, or even 
five, years ago.  There is an across-the-board 
drive toward the calculation of higher-order corrections
in perturbation theory,
as well as toward the perturbative-nonperturbative interface,
through resummed perturbation theory and related methods.
The growing awareness of, and capability for, spin studies
has reopened this field to perturbative QCD.  Under the
stimulus of new results, and projected experiments, 
theorists are reviving old ideas, inventing new directions,
and, in general, struggling to keep up!

In the remainder of this summary, I can give only selected
illustrations of these trends.  It will hardly be possible to
do justice even to the few topics I have space to discuss.
More information may, of course, be found from the plenary
talks at this conference, and from parallel sessions, and
almost all the references will be to these.

That understood, I shall begin with a few observations
on the excess of events at high $Q^2$ and $x$ reported
by H1 and ZEUS, which lead into an update on
perturbative QCD in DIS and beyond \cite{wg1,wg3}.
I will then turn to a review of generalizations of
the factorization formalism that underlies perturbative
QCD \cite{wg5}, from unpolarized to polarized scattering \cite{wg4} 
and to diffraction \cite{wg2}.  
I'll also say a few
words about openings to
nonperturbative QCD.  Finally, I will
conclude with some thoughts on where we are now,
and where the field is going.

\section{Structure Functions and the Excess at High $Q^2$}

The DIS cross sections $\sigma^\pm$ for ${\rm e}^\pm-{\rm p}$ scattering
is conventionally presented in terms of the structure functions
$F_2(x,Q^2)$, $F_3(x,Q^2)$ and $F_L(x,Q^2)$, 
\beqa
{d\sigma^\pm\over dxdy} &= & {2\pi\alpha^2 s\over Q^4}\, 
\left[ \left(1+(1-y)^2\right)F_2(x,Q^2)-y^2F_L(x,Q^2)\right.
\nonumber\\ 
&\ & \hbox{\hskip 0.5 true in}  \left.  \mp\left(1-(1-y)^2\right)F_3(x,Q^2)
\right]\, ,
\label{ftwotosigma}
\eeqa
where as usual $x=Q^2/2p\cdot q$ is the scaling variable with $q^2=-Q^2$,
and $y=Q^2/xs$ is the fractional leptonic energy loss in
the proton rest frame.  
Surely the most widely-recognized 
result from HERA this year is the excess of events
found at high $Q^2$ and $x$ (equivalently, high $y$ and $x$),
compared to the ``standard model", based on NLO QCD in DIS,
as seen by ZEUS \cite{ZEUSexpaper} and H1 \cite{H1expaper}.
Because the speculation engendered by these reports goes
far beyond the QCD that goes into them, I will discuss
their possible interpretation separately in this section,
and only then return to QCD, and the standard model with which they
have been compared.

It's an unexpected pleasure to comment on the
large-$Q^2$ excess, although it is hardly possible 
to review the theoretical studies that have sprouted like so many
flowers after a spring rain \cite{Dokplen}.
Surely it is, ``about time",  for a new
phenomenon, as it was so aptly observed.  
We shall have to wait and see, however, whether
the time has truly come.

\begin{figure}
\vskip 8.0 true in
\includegraphics{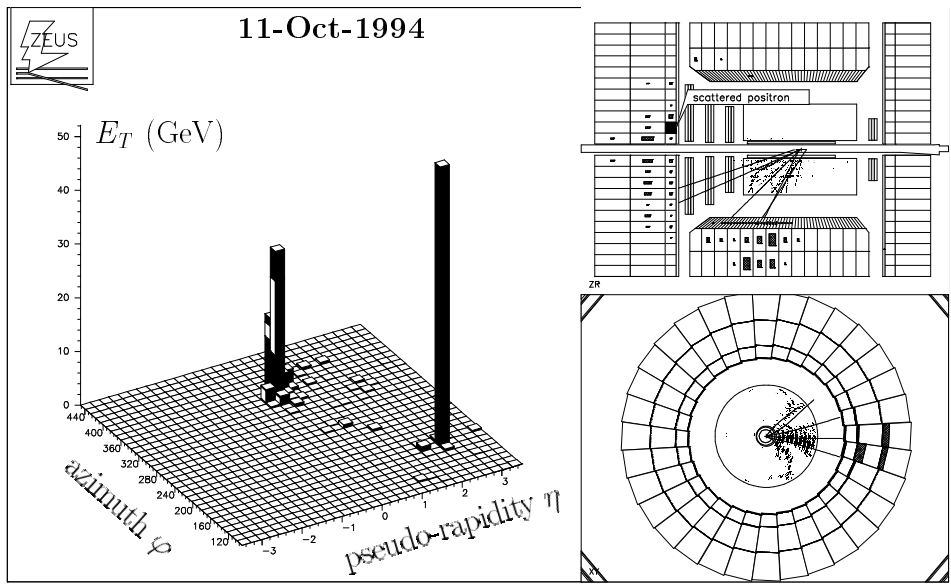}
\includegraphics{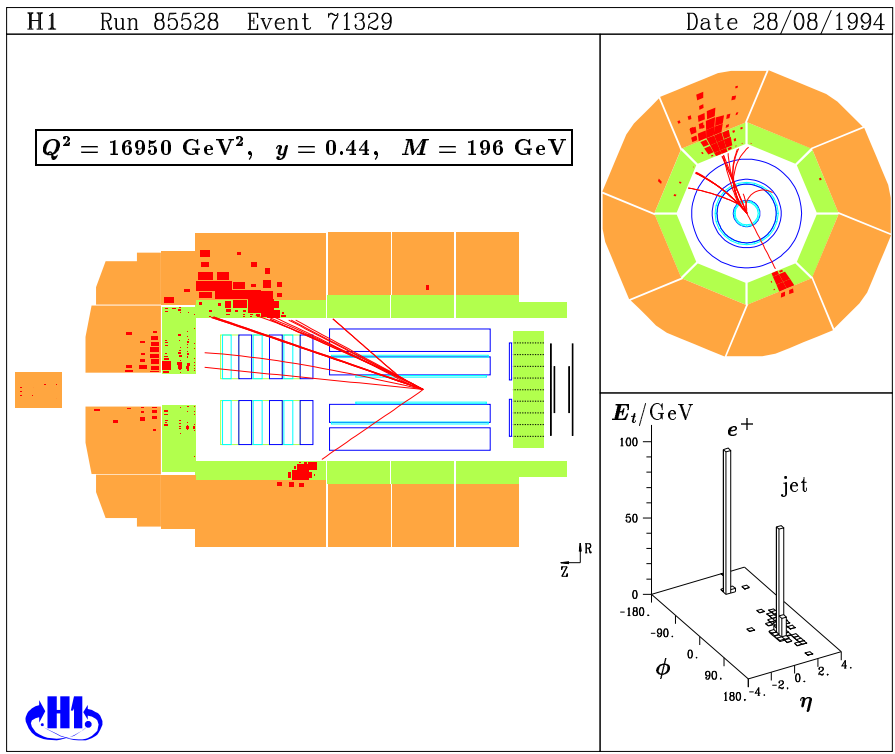}
\caption{Neutral current events in the highest-$Q^2$ sample,
as presented by  ZEUS {\protect\cite{ZEUSexpaper}} (top) and  H1
{\protect\cite{H1expaper}} (bottom).}
\label{Qevents}
\end{figure}

\begin{figure}
\vskip 3.0 true in
\includegraphics{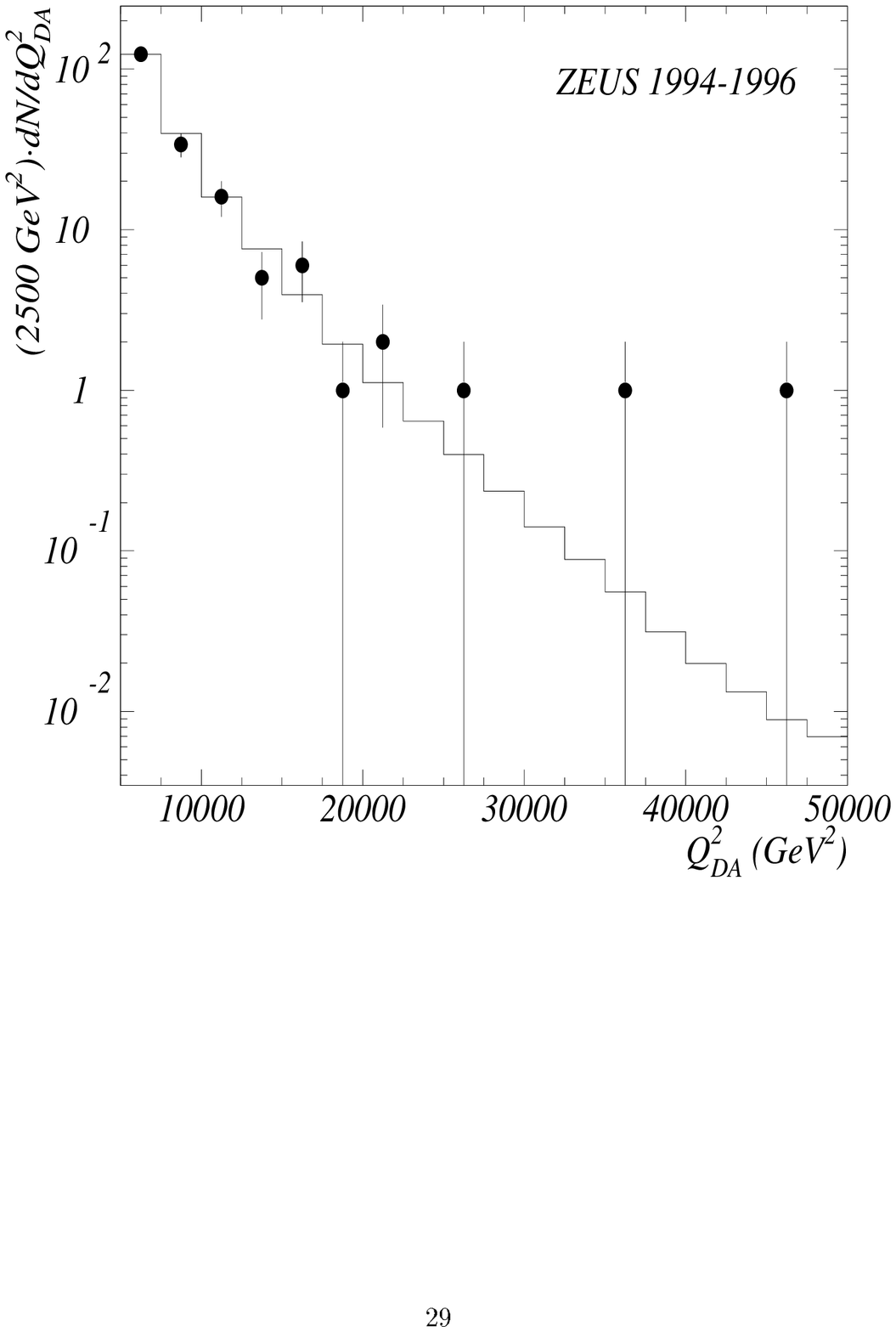}
\includegraphics{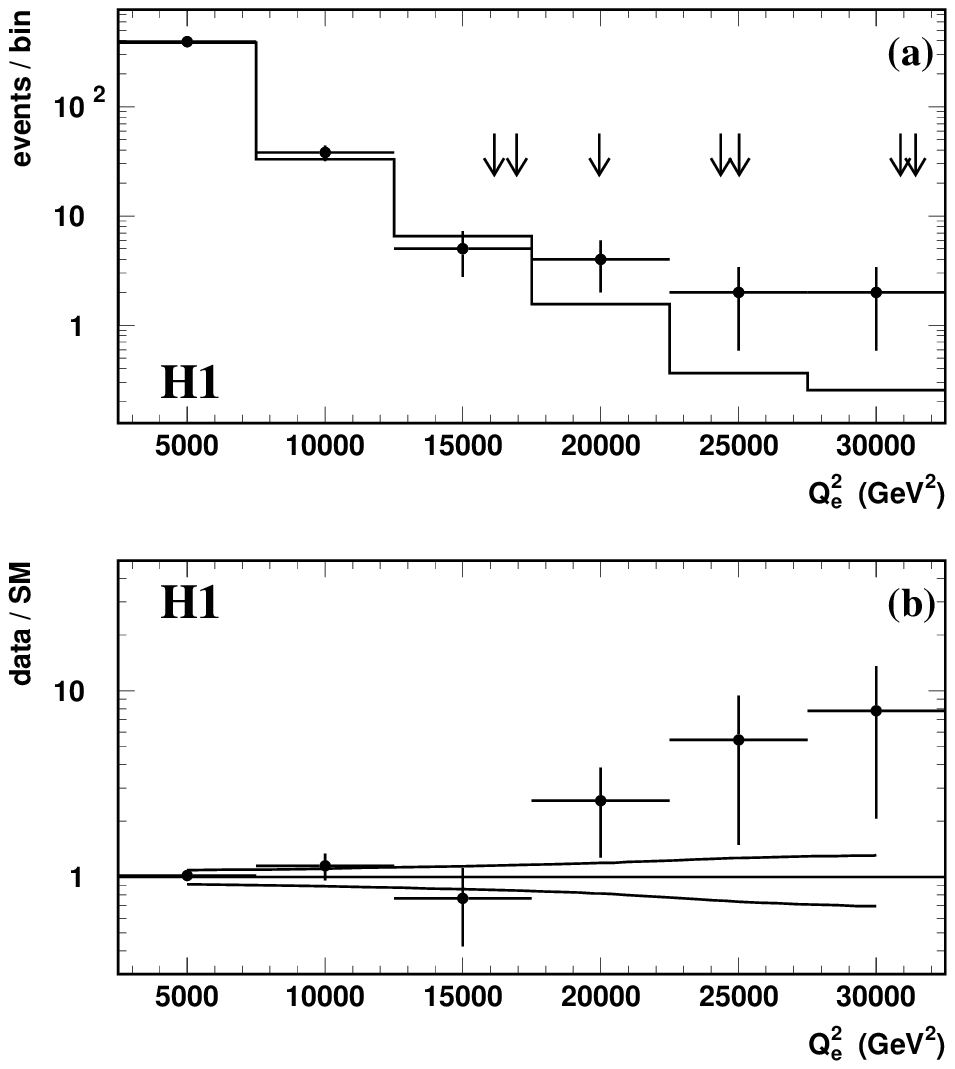}
\caption{$Q^2$ dependences of events seen at high $Q^2$,
compared to standard QCD predictions (histograms),
from  ZEUS {\protect\cite{ZEUSexpaper}} (left) and  H1
{\protect\cite{H1expaper}} (right). Subscripts DA (``double angle")
and {\it e} (``electron") refer to experimental
methods used in determining $Q^2$.}
\label{excess}
\end{figure}

The events themselves are beautiful examples of the
capabilities of the HERA detectors (Fig.\ \ref{Qevents}),
and of the imprint on hadronic final states of
momentum transfers at short distances.  The excess of
events over the presently-available ``standard model" 
is shown in terms of $Q^2$  in Fig.\ \ref{excess}.  

The most popular interpretations described at the workshop
rely either on a ``conservative" approach, of descriptions by
contact terms \cite{Zeppcontact}, 
or on more ambitious models with new particles,
leptoquarks, possessing both baryon and lepton quantum numbers 
\cite{Bleumlepto,Lola}. The most promising of these models
are based on supersymmetry, albeit in a somewhat 
variant form \cite{Lola}.

Contact terms are contributions to an
effective Lagrangian, which describes the
feed-down of very massive degrees of freedom to the standard model,
\beq
{\cal L}_{\rm contact}
=
\sum_{\stackrel{i,j=L,R}{q=u,d}}\, {\eta_{ij} \over \Lambda_{ij}^2}\;
{\bar e}_i\gamma_\mu e_i\; {\bar q}_j\gamma^\mu q_j\, ,
\label{contact}
\eeq
with $L$ and $R$ referring to left- and right-helicities.
These new terms are in direct analogy to the low-energy
four-fermion description of the weak interactions.
The parameters $\Lambda_{ij}$ are new scales, most likely associated
with heavy particle masses, typically in the TeV range.  Generally,
contact terms are not consistent with a ``bump" in the mass distribution
of the events, as marginally suggested by some of the H1 (but not ZEUS)
presentations of the data.

In leptoquark descriptions, the events are due to 
one (or more) new states in the actual mass range of the
events themselves, typically just above 200 GeV.  
Vector leptoquarks have larger cross sections than scalar,
and can be excluded by results from the Tevatron \cite{Wang}.
Even for scalar leptoquarks, Tevatron bounds are strong.
SUSY models are favored on this basis, because they result in 
suppressed branching ratios
for the more readily observable leptonic decays.

Since the new particles are evidently produced singly rather than 
as particle-antiparticle
pairs, candidate SUSY models cannot have the symmetry (``R-parity")
built into many SUSY models precisely to 
forbid such single-superparticle production
and decay mechanisms, and related (but avoidable) problems with 
proton decay \cite{Tata}.  In any case, the relevant terms in the
Lagrangian are of the general form
\beq
{\cal L}_{\rm SUSY}=\sum_{ijk}\lambda'_{ijk}L_iQ_jD_k\, ,
\label{susy}
\eeq
which couples leptons (through the corresponding
``superfield") $L_i$, quarks through $Q_j$ 
and the new superpartner, scalar leptoquarks,
through $D_k$.
Their coupling strengths are measured by the
constants $\lambda_{ijk}$, which must be determined
from experiment.  Bounds from various other experiments,
including $e^-p$ runs at HERA, limit the likely terms
in Eq.\ (\ref{susy})
to one coupling $e^+$ with the $d$ quark and a 
new scalar ``squark"
$\tilde c$, a superpartner of the 
 charm quark \cite{Lola,Altar}.

The impression left by the workshop is that the
most common explanations of this excess {\sl barely}
escape many bounds derived from a wide
class of other experiments.  These events seem to have been born into a
hostile world.  Some bounds are from explicit searches
in related experiments, most notably the Tevatron \cite{Wang}, where lower
limits on masses at the 95\% confidence level fall just short
of the excess, in the 150-200 GeV$^2$ range for scalar
leptoquarks, and seem to rule out vector leptoquarks 
altogether in this mass range.  The Tevatron and LEPII
\cite{Kalinow} put lower limits on
``contact terms" -- signals of the exchanges of very heavy particles
-- in the few TeV range, which promise to grow larger.  
At the same time, very
different, low energy precision experiments are equally,
perhaps even more, restrictive.  Of particular interest are
the atomic parity-violation experiments, which, by a special
serendipity, published new benchmarks of sensitivity 
within the few weeks prior to the 
conference \cite{Atom}.  Also striking are limits from
searches for rare $K$ decays and double beta decays \cite{Lola}.  

Depending on the scenario chosen to account for the 
excess, these already-existing experiments closely circumscribe
the range of parameters.  
This may be a good sign or bad, depending on what happens.
What is exciting is that something must happen.  H1 and ZEUS
have already begun an anticipated doubling of statistics
by the year's end.  Workers at the Tevatron advise us to
``stay tuned".  
If one of the favored explanations is correct,
it is probable that signals will show up elsewhere
soon, perhaps in events with pairs of high-$E_T$ leptons and jets 
at the Tevatron, perhaps in rare $K$ decays, perhaps in both.  

Other, more and less conventional, explanations were
also discussed, including a possible relation to
the much debated high-$E_T$ excess seen by CDF but (probably)
not by D0.  Such a connection is possible, although it might
be expected to be larger, not smaller, in hadronic
collisions \cite{Altar}.  Considered in isolation,
it seems just possible that the excess is ``simply" the sign of standard
parton distributions that are a bit larger at $x\sim 1$ than
in the present global fits \cite{KuhlTung}.  Exotic color
states were also suggested \cite{White}.
The adequacy of the ``standard" calculation for the
high-$x$ region might be reexamined as well.  One thing to
keep in mind is that any increase in the
QCD prediction would chip away at what is still a relatively
small statistical discrepancy between theory and experiment, and 
would make a (perhaps disappointing) explanation in terms
of a fluctuation more probable.
Certainly, we were left with much to think over.

\section{Perturbative QCD: DIS and Beyond, 1997}

Now it's time to return to QCD, the primary
theme of the conference. 
I will begin with a brief review of the theory that
underlies the experiments described here, 
and then discuss results 
on the fundamental objects in
perturbative QCD, structure functions and jet cross sections,
along with parton distributions and the
``BFKL" program.
Most of the essential features of this wide
range of topics must, of course,  be found in the
individual contributions to the workshop.

\subsection{The Basics}

The characteristic property of QCD is its asymptotic freedom,
according to which the coupling $\alpha_s(\mu^2)$
becomes weaker as it is probed 
at shorter distances $1/\mu$.  This extraordinary feature is
exploited by identifying infrared (IR) safe quantities,
cross sections (or other observables) which can be expanded
as power series in $\alpha_s$ in terms of coefficients 
that are IR finite, and more generally independent of
the light mass scales of the theory: $\Lambda_{\rm QCD}$
and the light quark and (vanishing) gluon masses.
If an IR safe cross section depends 
on large scale $Q$ and dimensionless kinematic variables
$x$, it takes the schematic form
\beq
Q^2{\hat \sigma}(Q^2,x)=\sum_{n\ge 0}c_n(Q^2/\mu^2,x)\;
\alpha_s^n(\mu^2)\, ,
\label{irs}
\eeq
where the $c_n$ are finite functions, or more often
integrable distributions in the variable $x$.
Classic examples of IR safe quantities are
the total cross section for lepton pair annihilation
to hadrons, and various jet and event shape cross
sections in the same process.

The power of perturbative QCD (pQCD) for large classes of inclusive processes
comes from its factorization and evolution properties.  Factorization is
expressed for unpolarized DIS of lepton $\ell$
on hadron $h$ as
\beqa
Q^2\sigma_{\ell h}(q,p,m)
&=&
\sum_{\rm partons\, {\it i}}
\int_x^1 d\xi\; 
{\hat \sigma}_{\ell\, i}
\left({Q^2\over \xi p\cdot q},{Q^2\over\mu^2},\alpha_s(\mu^2)\right)\nonumber\\
&\ & \quad \times  f_{i/h}(\xi,\mu,m)\, ,
\label{disfact}
\eeqa
\smallskip
where $q$ is the momentum transfer, $q^2=-Q^2$, $m$ represents the 
light scales of the theory (including the target mass $m=\sqrt{p^2}$),
and the sum is over parton type $i$.  The separation
of the IR-safe partonic cross section $\hat\sigma_{\ell\, i}$ from
the nonperturbative, but universal, parton distribution $f_{i/h}$ 
requires a factorization scale, $\mu$, usually identified
with the renormalization scale at which the coupling
is evaluated.  The physical cross section, of course,
is independent of $\mu$.  Corrections to (\ref{disfact})
are {\it power suppressed}, by at least $1/\mu^2$, so
that when $\mu$ is chosen to be of order $Q$,
factorization in terms of parton distributions is a
very good approximation at large momentum transfer.  On the
other hand, as $Q^2$ decreases below the proton mass,
we may expect ``higher-twist" power corrections to come into
play.  

Each of the DIS structure functions $F_i$,  $i=1,2,3$ satisfies 
a factorized form like Eq.\ (\ref{disfact}).
The convolution in parton fractional momentum $\xi p$ 
is conveniently denoted as, for instance,
\beqa
F_2 = C_2\otimes f\, ,
\label{otimesfact}
\eeqa
with $C_2$ an IR safe ``coefficient function", analogous
to $\hat{\sigma}$ in Eq.\ (\ref{disfact}). 
Our freedom to choose $\mu$ in the factorized DIS cross section,
\beq
\mu{d\sigma\over d \mu}=0\, ,
\label{dsigzero}
\eeq
readily leads to the DGLAP evolution equations \cite{DGLAP}, by separation
of variables,
\beqa
\mu{d{\hat\sigma}\over d\mu}&=&-{\hat\sigma}\otimes P(\alpha_s)\, ,
\nonumber\\
\mu{df\over d \mu}&=& P(\alpha_s)\otimes f\, ,
\label{dglap}
\eeqa
where the $P$'s are the familiar evolution kernels of QCD. 
They are perturbatively calculable precisely because $\hat\sigma$ is. 
Like the factorization formula, the evolution equations are
valid up to corrections suppressed by $1/\mu^2$.

Written out in terms of DGLAP kernels $P_{ij}$,
the evolution equations are
\begin{equation}
\mu^2{\partial \over \partial\mu^2}
\left(
\begin{array}{c}
q(x)\\G(x)
\end{array}
\right)
=
{\alpha_s(\mu^2) \over 2\pi}\; 
\left(
\begin{array}{cc}
P_{qq} & P_{qg}\\
P_{gq} & P_{gg}
\end{array}
\right)
\otimes
\left(
\begin{array}{c}
q(x)\\G(x)
\end{array}
\right)
\label{sevol}
\end{equation}
for singlet distributions, and
\beq
\mu^2{\partial \over \partial\mu^2}\; q_{\rm NS}
=
{\alpha_s(\mu^2) \over 2\pi}\; P_{qq}\;
q_{\rm NS}
\label{nsevol}
\eeq
for nonsinglet.  These, along with a set of boundary conditions, 
are used to predict parton
distributions and structure functions for previously unmeasured
values of $Q^2$.  (We should note that DGLAP
evolution alone does not allow us to ``evolve" in $x$; more
on this below.)  

A result of this past year deeply rooted in 
DGLAP evolution is the high-statistics determination of
$\alpha_s$, reported by CCFR through analysis of the
evolution of $F_2$ and $F_3$ \cite{ccfralphas},
using the proportionality of the logarithmic
derivative with respect to $Q^2$ of these
structure functions to $\alpha_s(Q^2)$. With this result,
$\alpha_s(M_Z^2)=0.119\pm.002$, measurements
of the strong coupling from relatively low-energy DIS experiments
have come (predominantly, if not completely \cite{Schmelling}) into agreement with
values found from ${\rm e}^+{\rm e}^-$ annihilation at the Z mass.
This has put to rest some of the speculation based on
discrepancies between the two ranges of energy.  Not all
DIS estimates give such a high value, however, although 
scaling violation in $F_3$ appears to be the most precise.

\subsection{Structure Function Measurements}

ZEUS and H1 presented data, and new analyses of data,
from 1994 through 1996, mapping
out the behavior of  $F_2$ over an increasing
range of $x$ and $Q^2$, with shrinking errors 
\cite{wg1,Meyer,Surrow}.  The primary emphasis was on the very high-$Q^2$ 
and very low-$Q^2$ results.  These were complemented by
the final NMC results for Deuterium, at large $x$ and moderate $Q$
\cite{nmc}, and those from CCFR \cite{ccfr}.

\begin{figure}
\centerline{\epsfig{file=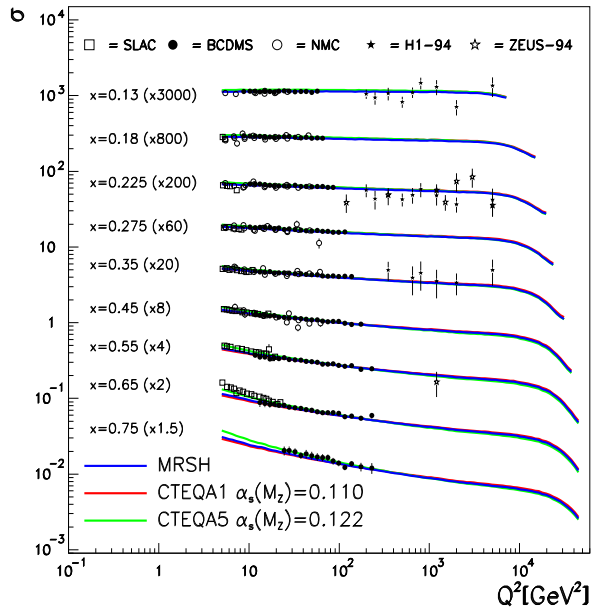,height=3.5in,width=3.5in}}
\vspace{10pt}
\caption{(a) H1 fit to high-$Q^2$
neutral current cross section approximately normalized to $F_2$ at low energies
{\protect\cite{H1plen}}.}
\label{sigtild}
\end{figure}

The high-$Q^2$ analysis at HERA became
possible as the integrated luminosity
passed 20 pb$^{-1}$ by the end of 1996.  We have already discussed
the observed excess of events in that region.  A feeling for the 
nature of the DGLAP-based evolution of the neutral current
cross section
to large $Q^2$ is given by Fig.\ \ref{sigtild},
which plots the cross section of Eq.\ (\ref{ftwotosigma})
scaled by the coefficient of $F_2$. 
 On the basis of analyses such as these, H1 and
ZEUS have estimated uncertainties in the NLO predictions at high $Q^2$ of order
ten percent \cite{Botje}.

\begin{figure}[t]
\centerline{\epsfig{file=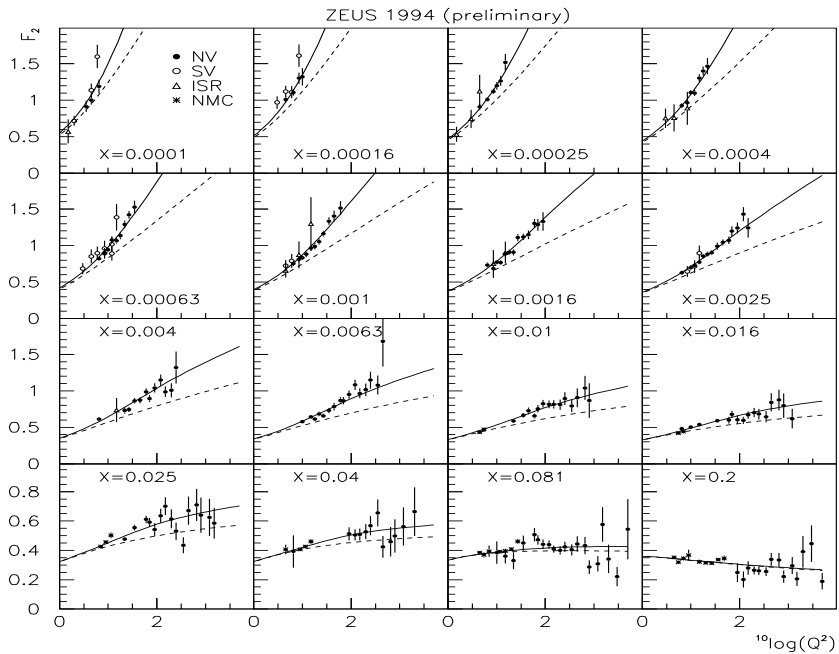,height=4.0in,width=5.5in}}
%\special{psfile=sfig4.eps hscale=100 vscale=100 hoffset=-100 voffset=-200}
\caption{$Q$-dependence of $F_2$ for representative $x$ at low $Q$
{\protect\cite{Botje}}.  The solid curve is a fit, the dashed curve
omits the charm contribution.}
\label{xfixed}
\end{figure}

\begin{figure}[t]
\centerline{\epsfig{file=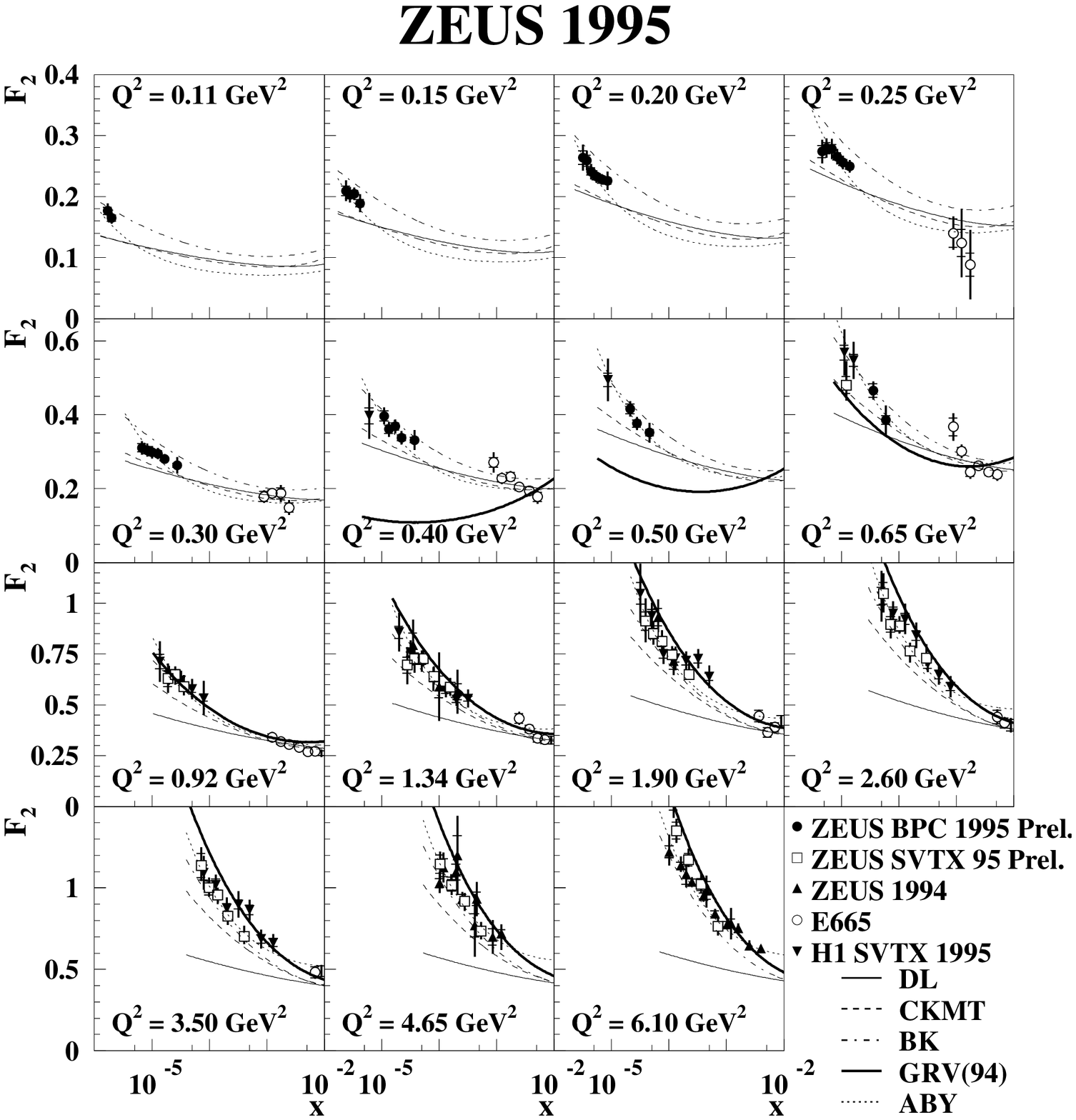,height=4.5in,width=5.5in}}
%\special{psfile=sfig5.eps hscale=60 vscale=60 hoffset=-200 voffset=-200}
\caption{$x$-depndence of $F_2$ for low $Q^2$ {\protect\cite{Surrow}}.}  
\label{Qfixed}
\end{figure}

New low $x$ and $Q^2$
results from HERA were made possible by runs with detector modifications
and/or shifts in the event vertex, which extend coverage of the
existing detectors to more forward regions \cite{Meyer,Surrow}.
Figs.\ \ref{xfixed} and \ref{Qfixed} show, respectively,
trends of the data  with $Q$ for representative values of
$x$ and vice-versa.


As Fig.\ \ref{xfixed} shows, charm production plays a large role
in the total cross section for small $x$ even for relatively
small $Q^2$, comparable to the charm mass \cite{roldan,Zomer1}.  This is understandable;
the relevant variable is now $W^2\sim Q^2/x$ at small $x$, and
we can be far above the charm threshold even for $Q<m_c$, if
$x$ is small enough.  Indeed, charm production is an important
effect even for photoproduction, where $Q^2=0$ \cite{charmphoto}.

The special interest of low $Q$ comes in part from the 
very success of DGLAP evolution in describing 
$F_2$ to surprisingly low values of $Q^2$, even below 1 GeV$^2$.
The GRV parton distributions \cite{GRV} are based on ``valence-like" 
boundary conditions in this range, and give a reasonable
description of $F_2$ there.  Below $Q^2\sim1$ GeV$^2$, deviations begin to
show up, in which the cross section grows less rapidly toward
small $x$, corresponding to a slow-down of evolution compared
to perturbative predictions.  The picture here remains a bit
cloudy, with various explanations, fits \cite{Haidt}
and models offered, often inspired by pre-QCD 
phenomenology, such as vector meson
dominance \cite{BK} and Regge theory \cite{DL}, alongside efforts at quantitative
prediction based on the perturbative BFKL formalism \cite{KMS}, which
itself has one foot in QCD perturbation theory and one in Regge theory.  
The transition region between perturbative and nonperturbative
degrees of freedom is not wide, at least as measured in GeV,
but no single model can yet bridge it.  We must hope for theoretical
developments in this direction, a question to which we shall
return.

In terms of pQCD itself, we can identify two
generic corrections for low $x$ and $Q$ not included in the 
DGLAP equation at NLO, which warn us of the
transition region.  These are: logarithms of
$x$ at higher orders in the splitting kernel, and
power corrections that are order $1/\mu$ with
$\mu\sim Q$ the factorization scale.  It is important
to emphasize that these corrections are different. 
In particular, the former are consistent with leading
twist factorization, the latter require (at the least)
generalized factorizations.
The BFKL equation deals with logarithms of $x$ at leading power.
Let's review how it arises in DIS.

\subsection{The BFKL Program: Theory and Experiment}

The well-known BFKL equation \cite{bfkl}, applied to DIS, summarizes 
leading logarithms in $x$.  Let me reemphasize that
the BFKL equation is fully consistent with DGLAP evolution,
and constitutes a reorganization  of information in the
DGLAP kernels $P$. Like the DGLAP equation, it may be derived
from a factorization of DIS structure functions, but now
one that is accurate only to the level
$1/\ln(x)$,
\beqa
F(x,Q^2)&=& \int_x^1 {d\xi\over\xi}\; 
C\bigg({x\over\xi},{Q^2\over \mu^2}\bigg)
G(\xi,Q^2)+{\cal O}\left ({1/Q^2}\right )
\nonumber \\
&=& \int d^2k_T\; c\bigg({x\over\xi'},Q,k_T\bigg)\psi(\xi',k_T)\
+{\cal O}\left({1/\ln (1/x)}\right )\, .
\label{bfklfact}
\eeqa
In the second form, the wave function $\psi$ is in a $k_T$ convolution
with the coefficient function, while the fractional momentum $\xi'$
plays the role of a factorization scale.  The relation of
$\psi$ to the gluon distribution $G$ is
\beq
G(\xi,Q^2)=\int^Qd^2k_T\; \psi(\xi,k_T)\, .
\label{gpsi}
\eeq
Just as the first factorization form leads to DGLAP evolution by invoking the
independence of $F$ on $\mu$, so the independence of $F$ on $\xi'$
leads to another evolution equation, analogous to Eq.\ (\ref{dglap}),
but now with a convolution in transverse momentum rather than
$\xi'$.  This is the BFKL equation \cite{bfkl},
\beq
\xi{d\psi(\xi,k_T)\over d\xi}=\int d^2k'_T\; {\cal K}(k_T-k'_T)\psi(\xi,k'_T)\, .
\label{bfkl}
\eeq
The kernel $\cal K$ turns out to be independent of $\xi$ in this
approximation, although by dimensional analysis it could
have depended upon it.

 Further insight into the
significance of the BFKL equation
can be found in the
very general ``Wilsonian" treatment \cite{Kovner}
of the underlying factorization, by
separating the dynamics into independent
Hamiltonians, appropriate to the
opposite-moving external particles.
I think we can anticipate progress
in interpreting other factorizations
from this approach in the coming years.

Solutions to the BFKL equation may be found by substituting 
trial solutions in the form of powers of $\xi$ and $k_T$.
These solutions take a continuous range of powers of $\xi$,
of which the most singular is the famous BFKL result,
\beq
\xi\psi(\xi,k_T)\sim\xi^{-4N\ln 2(\alpha_s/\pi)}\, .
\label{bfklpom}
\eeq
The scale, and therefore the size, of $\alpha_s$ is undetermined in 
this leading logarithm
formalism.   In addition, the full
description of near-forward scattering at next-to-leading logarithm
in $x$ requires three-gluon exchange contributions, which 
are not included in the BFKL equation at all. 
These cautions aside, a provocative treatment of
nonleading logarithms in solutions to the BFKL
equations based on analogy to DGLAP evolution was
presented \cite{ballforte}.

From the DGLAP point of view, the BFKL
equation sums up those contributions
to the gluon-gluon splitting function
that are leading at each order in $\alpha_s$, of
the form $\alpha_s^n\; [\ln^{n-1}(1/x)]/x$.
It is by now an old story, however, that the growth in
$F_2$ at low $x$ should not be interpreted 
as a direct observation of unadorned
BFKL dynamics. NLO DGLAP evolution seems quite up
to the task, and the higher orders
seem not to be needed.  

In any case, the solution Eq.\ (\ref{bfklpom}) 
cannot strictly speaking be exact, because
it predicts a gluon distribution
that increases as a power of $x$, and therefore
a total cross section that increases as 
a power of $W^2\sim 1/x$, which would
eventually violate unitarity bounds.   
One way of understanding this behavior is
that power-like solutions do not build in
momentum conservation, which is only enforced
at the level of nonleading logarithms.

At the same time, when $x$ is small enough that
$\alpha_s\ln(1/x)\sim 1$, nonleading logarithms
must be taken into account, and have the effect
of moderating the growth of $G(x)$ with $1/x$.
This general picture is consistent with what
we have seen above, in Fig.\ \ref{Qfixed}, for 
instance, but there is no assurance that by
the time resumed BFKL logarithms
 become important for
structure functions, a perturbative
treatment is appropriate at all.  Nevertheless,
it is possible to construct models for $\psi(x,k_T)$,
based on the evolution equations we have encountered
above, and to derive in this way a fairly good
picture of low-$x$ and $Q^2$ evolution \cite{KMS}.

The search for BFKL phenomenology, that is,
for experiments whose interpretation
requires the resummation of
logarithms of $1/x$, has by and large shifted
from the structure functions to less
inclusive measurements, following general
lines introduced by Mueller and Navelet \cite{MueNav} in
hadron-hadron scattering.  The method 
is to identify events that can be
``tagged" by jets at wide rapidity separations.
If the transverse momentum of the jets is substantial,
there is little chance of a higher-$p_T$ jet
in between, and DGLAP evolution should be unimportant.
At the same time, if the partonic fractions of the two jets
are very different, perturbative BFKL dynamics ought to dominate.
At {\it very} high energy, the solution (\ref{bfklpom}) would 
describe such cross sections,
with $\ln(1/\xi)$ replaced by the rapidity difference between the two jets.
Energy conservation, however, makes it
impossible to have enough gluons emitted between
the tagged jets to reproduce the exponential
of the rapidity, and more sophisticated
numerical approaches appear to be necessary.
Among these are very new BFKL ``Monte Carlo" generators
\cite{Schmidt}.  The general
features of BFKL evolution have been built into
the event generator ARIADNE, at a somewhat less
formal level.  

\begin{figure}
\centerline{\epsfig{file=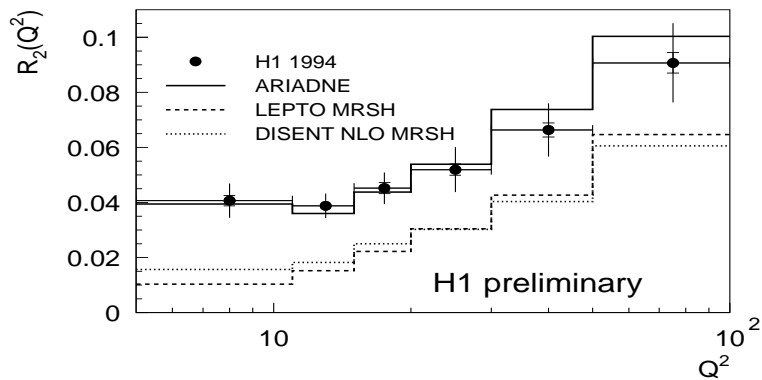,height=2.0in,width=4.0in}}
%\vskip 2.5 true in
%\special{psfile=sfig6.eps hscale=100 vscale=100 hoffset=0 voffset=0}
\caption{$R_2\equiv (N_{2\; {\rm jet}}/N_{\rm all})$ 
measured by H1 {\protect\cite{H1plen}}, compared
to NLO calculations and ARIADNE.}
\label{dijetdis}
\end{figure}

Studies searching for such BFKL effects have
been carried out at the Tevatron \cite{Jun}
in terms of angular correlations between the
tagged jets, without showing 
specifically BFKL behavior.
At the same time, studies of dijet
cross sections by ZEUS
\cite{Wolfle}
at large rapidity separations
are more encouraging, at least when ARIADNE
is compared to NLO calculations \cite{Zepp3}.  A
corresponding example from H1 is
shown in Fig\ \ref{dijetdis}.  The difference between the
two reports may be more apparent than real, however,  since
the ``theory" used in each case is different.

Finally, progress has been reported in the
extension of the BFKL equation.
The program of Fadin and Lipatov
toward the calculation
of the two-loop kernel for the BFKL equation, 
appears to be at the threshold of completion \cite{fadin}.
The capstone of this enterprise, the ``two-reggion-two-gluon vertex",
has been presented at this workshop for the first time.
The expression remains at this time a bit unwieldy to 
speculate on its quantitative consequences, although
expected signs of the running of the coupling are possible
to pick out. 
Alternative technical approaches, based on
helicity methods  are also being pursued \cite{ddel}.

\subsection{Direct Evolution for Structure Functions}

To incorporate the extra information  on evolution of
DIS structure functions
contained in the BFKL equation, while still keeping
logarithms of $Q$, it is
useful to {\sl demand} an evolution that is
independent of changes of 
factorization scheme, up to unincorporated
corrections in $\ln(1/x)$ as well as $\ln Q^2$.  The
latter are taken care of by DGLAP evolution.  The former
may be incorporated by studying the evolution of the
structure functions themselves \cite{Thorne}.
Combining factorization
\beq
F_2^{\rm NS}(Q)=C\left(\alpha_s(Q^2)\right)
\otimes
q_{\rm NS}(Q)
\label{nsfact}
\eeq
with the nonsinglet evolution equation (\ref{nsevol}), 
we readily derive an evolution equation for the nonsinglet
structure function itself,
\beq
{\partial \over \partial \ln Q^2}\; F^{\rm NS}(Q)
=
\left ({\partial \ln C\over \partial \ln Q^2}
+C\otimes P_{qq}\otimes C^{-1} \right)
\otimes
F^{\rm NS}(Q)
\equiv \hat{\Gamma}_{\rm NS}\otimes F^{\rm NS}\, ,
\label{nsFevol}
\eeq
in which $\hat{\Gamma}_{\rm NS}$ includes the evolution
dependence of the coefficient function through
the running coupling.  The extension to singlet
structure functions follows the same pattern.

This formulation makes possible a ``consistent", ``scheme-independent" 
treatment of logarithms of $x$, through the incorporation
of coefficient functions, along with evolution in $Q^2$.
As evolution proceeds, the input level of accuracy 
in $\ln x$ and $\alpha_s(\mu)$ are retained, by
avoiding scheme-dependent complications which
can arise from separating the coefficient functions
from the parton distributions.
This makes possible 
surprising improvements in the fit to the low-$x$ HERA data \cite{Thorne}.

Interestingly, the same general approach can be used a very high $x$,
to take into account logarithms of $1-x$ not included in NLO \cite{LS}.
In this case, the nonsinglet evolution equation alone is adequate,
as logarithms of $1-x$ do not involve flavor mixing.
The equation that takes into account these corrections  is exactly Eq.\ (\ref{nsFevol}), 
in terms of DIS scheme quark distributions and a resummed
coefficient function $C(x)$.  The logarithms of $1-x$ are
generated from moments,  $\tilde{C}(N)=\int_0^1
dxx^{N-1}C(x)$.  $\tilde{C}(N)$ is known in an explicit
form in which
all leading and next-to-leading logarithms are exponentiated \cite{expon1x}
\beq
 {\tilde C}(N,\alpha_s(Q))
=
\exp \left[ -C_F\int_0^1dz{z^{N-1}-1\over 1-z}\; 
\int_{(1-z)}^{1}{d\eta \over \eta}{\alpha_s(\eta Q^2)\over\pi}+\dots\right ]\, .
\label{CofNresum}
\eeq
Evolved according to Eq.\ (\ref{nsFevol}), the DIS scheme distributions, whose
sum is the structure function in this limit, will be enhanced
compared to the result found by evolving either $\overline{\rm MS}$
or DIS distributions according to NLO DGLAP evolution alone.

\subsection{Parton Distributions and Charm}

An analysis directly in terms of structure functions,
however useful for inclusive DIS, is limited to 
that process.  To treat jet cross sections, or heavy
quark production in DIS and other
processes, we need to retreat from the ascetic
evolution in terms of observables only, to parton distribution
functions. Then, the generic cross section is
of the form
\beq
\sigma_{hh'}= \sum_{partons\ ij}f_{i/h} \otimes f_{j/h'} \otimes \hat{\sigma}_{ij}\, ,
\eeq
where $\hat{\sigma}$ is an IRS distribution, calculable in pQCD.  

\begin{figure}[t]
\centerline{\epsfig{file=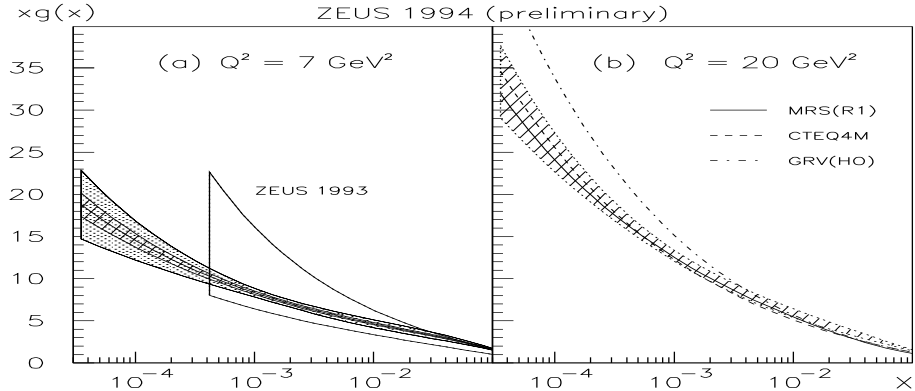,height=2.5in,width=5.5in}}
%\vskip 2.5 true in
%\special{psfile=sfig7.eps hscale=100 vscale=100 hoffset=0 voffset=0}
\caption{The gluon distribution fit of ZEUS, from {\protect\cite{Botje}}.}
\label{gluondist}
\end{figure}

\begin{figure}
\centerline{\epsfig{file=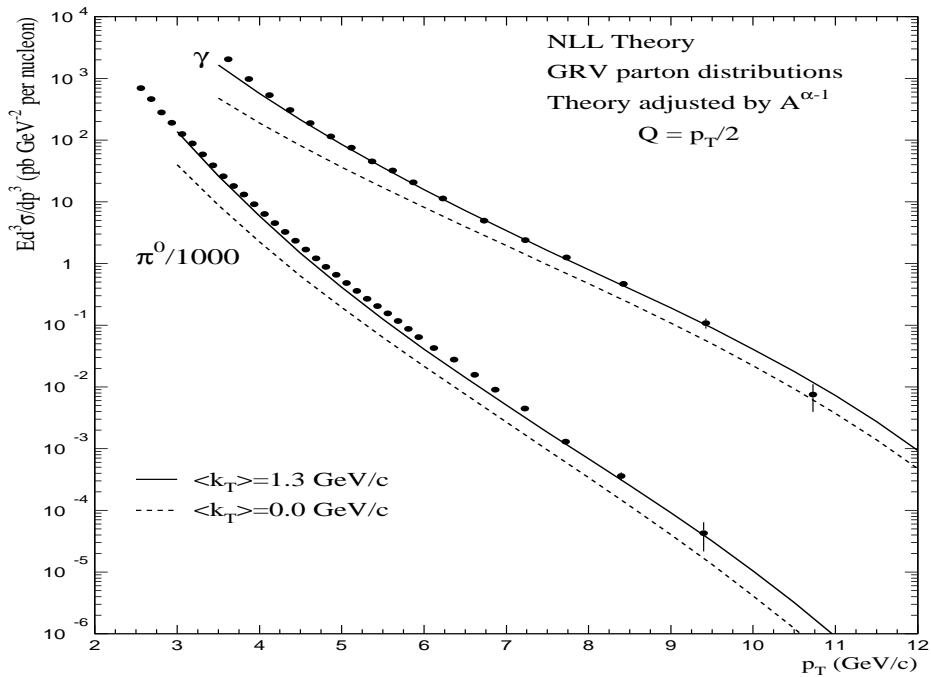,height=4.0in,width=5.0in}}
%\vskip 3.5 true in
%\special{psfile=sfig8.eps hscale=35 vscale=35 hoffset=0 voffset=0}
\caption{Direct photon and $\pi^0$ inclusive cross sections measured
by E706 {\protect\cite{Zielinski}}.}
\label{gammaonepi}
\end{figure}

The gluon distribution can be
determined by fitting $F_2$ at NLO \cite{Botje}, and is
also accessible at leading order in jet, heavy quark and direct photon
cross sections \cite{Zomer3,Mikunas,Huston,Kosower1}.
An example of the former is shown in Fig.\ \ref{gluondist} from \cite{Botje},
which exhibits a good agreement with results of
the global fitting programs of MRS \cite{mrsr1} and CTEQ \cite{cteq4m}.
The more exclusive determinations, such as those based on direct
photons, are appealing because the gluon distribution enters them
at leading order.  At the same time, higher-order corrections are
generally more difficult to control in these cases, and sometimes effects that
are formally nonleading twist, such as ``intrinsic" parton momenta,
are important \cite{Zielinski,Blair}.  Indeed intrinsic
transverse momenta of order 1GeV 
seem necessary to explain the comparison of NLO theory with experiment
for direct photon production measured by the fixed target
experiment E706, as illustrated by Fig.\ \ref{gammaonepi}. 
Non-NLO effects of this type should be anticipated in  any
single-particle inclusive or related measurement.

\begin{figure}[t]
\centerline{\epsfig{file=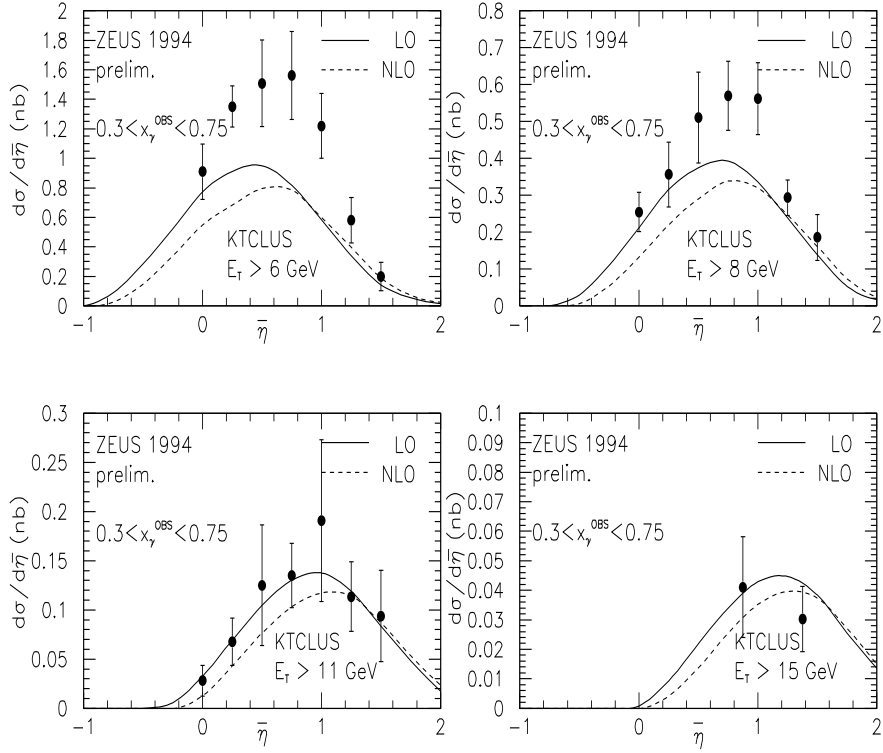,height=4.5in,width=5.5in}}
%\vskip 3.0 true in
%\special{psfile=sfig9.eps hscale=100 vscale=100 hoffset=0 voffset=0}
\caption{Cross sections for photoproduction of dijets
in various ranges of $x_\gamma$ {\protect\cite{Forplen}}.}
\label{photodij}
\end{figure}

Such considerations also arise in connection with
the program of measuring parton distributions for photons through
``resolved" photoproduction experiments.
Partons are associated with dijets, whose rapidities
and transverse momenta $\eta_i$ and $E_{Ti}$ are
related to an equivalent partonic fraction,
\beq
x_{\rm \gamma}={1\over 2yE_e}\left(E_{T1}{\rm e}^{-\eta_1}+E_{T2}{\rm e}^{-\eta_2}\right)\, .
\label{equivx}
\eeq
 The observation of Rutherford-like
cross sections for ``resolved-enriched" $x_\gamma<1$
in jet photoproduction has demonstrated the 
feasibility of such determinations \cite{Saunders,Ebert3}, and 
the characteristic evolution of the quark distribution has
been observed.  
Recent measurements, however,
show enhancements in the resolved region 
(Fig.\ \ref{photodij})
which seem to be due
to multiple interactions \cite{Forplen,Lamoureux}.  Such
interactions are also ``higher twist" -- formally power-suppressed in $p_T$.
Again, a fuller understanding of such non-NLO effects will be
necessary in these cases.

A better-understood, but still challenging issue is the role
of charm in DIS.  We have already seen that the charm
structure function $F_2$ is a substantial part of the total (see Fig.\ \ref{xfixed}).
In response to this new data, 1997 has become
the year of quark masses in parton distributions 
\cite{tunghq,robertshq,lainewdist,olnesshq}.  
The problem here is to develop a treatment that
is at once self-consistent, and accurate, when
parton distributions are evolved through $\mu\sim m_c$,
where $m_c$ is the charmed (or other) quark mass,
$m_c\gg \Lambda_{\rm QCD}$.  

We note first that since $m_c\gg \Lambda_{\rm QCD}$,
it is perfectly 
acceptable to treat the $c$ quark as perturbatively
generated, and to work throughout in a ``three-flavor"
scheme, for which the {\it only} quark distributions
are $u(x)$, $d(x)$ and $s(x)$.  This is a 
self-consistent approach up to order $\Lambda_{\rm QCD}/m_c$, but it suffers from
(at least one) serious drawback.  For scales $Q\gg m_c$, 
we must be willing and able to compute very high-order
diagrams in which $c$-quark pairs are produced, or
be prepared to lose lots of terms that behave as $\alpha_s^m(Q^2)\ln^m(Q/m_c)$  
in the coefficient functions in Eq.\ (\ref{otimesfact}).
This problem becomes worse and worse at small $x$, since
the phase space available for $c$-pair production is measured
not by $Q^2$, but by $W^2=Q^2(1-x)/x$.  

The provisional solution to these problems adopted by
the global fits until recently was to take the $c$
quark into account
by ignoring it for $\mu<m_c$, and treating it as
massless for $\mu>m_c$, switching from a three-
to a four-quark fit at that point. Clearly, this
is only a rough picture of the relevant physics. 

\begin{figure}[t]
\centerline{\epsfig{file=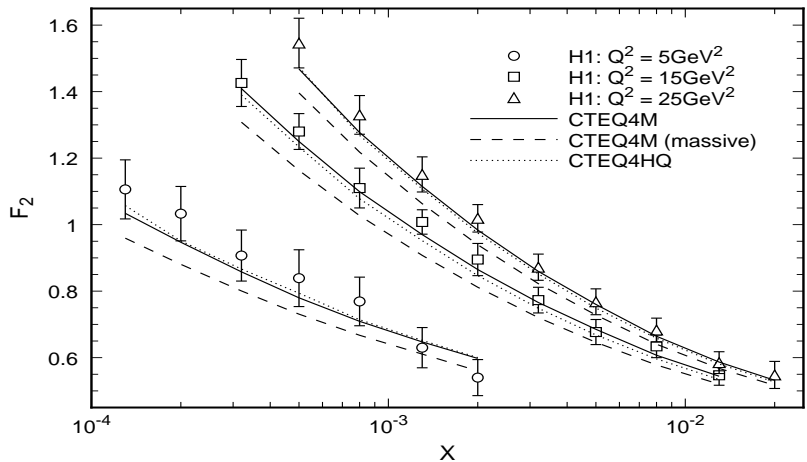,height=2.5in,width=4.5in}}
%\vskip 2.5 true in
%\special{psfile=sfig10.eps hscale=100 vscale=100 hoffset=0 voffset=0}
\caption{Fits of H1 measurement of $F_2$ using the (old, massless)
CTEQ4M fits, with and without quark masses in the coefficient function,
and the variable flavor fit CTEQHQ.}
\label{varflav}
\end{figure}

A much more appealing approach is to absorb 
powers of $\ln(Q/m_c)$ into the evolution of the
$c$ quark distribution,
\beq
{\partial \over \partial \ln \mu^2}f_c(\mu)={\alpha_s\over2\pi}\;
P_{qq}\otimes f_c(\mu)\, .
\label{fcevol}
\eeq
As described in \cite{tunghq,olnesshq}, the kernel $P$ is the 
standard $\overline{\rm MS}$ kernel, a result that
follows by showing that it is possible to construct
a factorized expression for the cross section
\beq
\sigma(Q,m_c)\sim \hat{\sigma}(Q/\mu,m_c/\mu)\otimes f_c(\mu)\, ,
\label{cfact}
\eeq
in which the hard-scattering function obeys
\beq
\hat{\sigma}(Q/\mu,m_c/\mu)=\hat{\sigma}(Q/\mu,0)+O(m_c/\mu)\, ,
\label{mctozero}
\eeq
so that, as above,
\beq
{\partial \over \partial \ln \mu^2}\hat{\sigma}=-\hat{\sigma}\otimes 
P^{\overline{\rm MS}}\, .
\eeq
The result of a new CTEQ fit \cite{tunghq,lainewdist}, employing this scheme, is shown in
Fig.\ \ref{varflav}.
These, and related \cite{robertshq}, ``variable flavor schemes"
seem to be demanded for a full description of the small-$x$ 
DIS data, and correspondingly for applications in hadron-hadron
scattering.  This year has seen truly significant progress
in this direction.

\subsection{QCD at NLO and NNLO: How Much Can We Expect?}

QCD predictions based on factorization,
next-to-leading order calculations and parton
distributions run the gamut from spectacular
success to egregious failure.  
An example of recent success is found in
jet cross sections at the Tevatron, at least
up to transverse momenta of 200 GeV \cite{Hirosky}, while
a particularly interesting recent failure is
in the ratio of $W$ cross sections with and
without a single jet \cite{JoffeM}.  

All other things being equal, NLO predictions
work best when all
scales are comparable.
Thus, top production
is easier to handle than bottom production at
the Tevatron, while DIS at $x$ of order unity
is better described at NLO than when
$x\rightarrow 0$.  Similarly, steeply falling
cross sections are sensitive to nonperturbative
$k_T$ and/or transverse momenta from
high orders (see Fig.\ \ref{gammaonepi} above). 
Finally, any edge of phase space,
even $p_T^{\rm max}$, is dangerous.  For example, the
default power suppression in DIS is $1/[(1-x)Q^2]$,
and similarly for fragmentation,
an effect that may be at the basis of the 
rise in octet contributions to $\psi$ 
production at $z\rightarrow 1$ \cite{Beneke}.

We should note, however,
that our expectations for sucess have become
more demanding, and our recognition of failure
more ready.   This is due in large part to 
the accumulation of NLO calculations \cite{Ellisbook},
first for electroweak processes, such as 
Drell-Yan and DIS, for direct photon,
heavy quark and jet production in hadronic
collisions \cite{Kilgore}, and for jet production in DIS \cite{Zepp3}
and ${\rm e}^+{\rm e}^-$ annihilation \cite{DixonSigner}.
Certainly, next-to-next-to-leading-order calculations
can lead to further improvements, although
except when all relevant scales are very large,
power corrections can be competitive with
NNLO.  

Prospects for improved partonic calculations
were extensively discussed at the workshop.
Certain very basic quantities, such as
the QCD beta function in certain gauges,
and the first moment of the polarized-DIS
structure function $g_1(x)$ already can be
computed to four (!) loops \cite{Rittenberg}.  More generally,
the use of still-new techniques of supersymmetry
and strings are pointing the way toward two,
and even higher, loop amplitudes in QCD.  The
status of these calculations is, perhaps, similar
to that of NLO calculations in the late 70's.  Time
will tell if the technology of two loops can
reach the sophistication, and phenomenological 
impact, of the NLO calculations that grew out
of the pioneering efforts of that time.  

The string
approach, employed originally to calculate
massless one-loop QCD helicity amplitudes \cite{Kosower5},
has been largely superseded by methods based on judicious
use of SUSY, for instance, by expressing QCD amplitudes as 
linear combinations of amplitudes in
theories with four  
supersymmetries (``$N=4$") and one supersymmetry (``$N=1$"), combined
with clever decompositions of color amplitudes.  The
overall program has been labelled ``total quantum number management".
The most striking simplification \cite{Bern}, however, is based on a
very old idea, 
that loop amplitudes may be constructed iteratively using
 ``cutting rules", which specify the discontinuities
of Feynman diagrams as products of amplitudes and
complex conjugates found simply by dividing the diagrams in
two.  At the same time, progress was reported on deriving
-- in principle -- QCD amplitudes at {\it arbitrary}
loops, by a new approach to string techniques \cite{Magnea}.
The practicality of these new techniques,  of  course,
is undemonstrated, but their potential is great. 

\section{Other Factorizations: New Densities, New Evolutions}

The value of the factorization of long- from short-distance 
dynamics goes far beyond unpolarized structure functions
and related hard-scattering cross sections.  This year has
shown, not only applications to polarized structure functions,
but also the extensions of the general method of factorization
to less inclusive final states.  The value of an extension
of the factorization program depends upon the experimental,
as well as theoretical accessibility of the relevant
cross section.  

\subsection{The Sibling Distributions: Polarized DIS}

Inclusive polarized parton distributions 
are at the next level of hadron structure from
the unpolarized distributions discussed above. 
They are ``siblings" of unpolarized distributions,
in the sense that they were part of the 
 parton model analysis of DIS.

Their analysis begins with the hadronic tensor
for polarized DIS,
\beqa
{d^2\sigma\over d\Omega dE}
&=&
{\alpha_{\rm EM}^2 \over 2mQ^4}{E_e\over E'_e}
L^{\mu\nu}W_{\mu\nu}
\nonumber\\
W_{\mu\nu}&=&W_{\mu\nu}^{\rm unpol} + {i\over E_e-E'_e}\epsilon_{\mu\nu\lambda\sigma}q^\lambda s^\sigma\; g_1(x,Q)
\nonumber\\
&\ & + {i\over (E_e-E'_e)^2}\epsilon_{\mu\nu\lambda\sigma}q^\lambda\left[p\cdot q s^\sigma
-
s\cdot q p^\sigma\right ] g_2(x,Q)\, ,
\label{Wpol}
\eeqa
with $s$ the nucleon spin.  
From the polarized structure functions $g_1$ and $g_2$, we abstract polarized
parton distribution functions, for instance
\beq
g_1(x,Q) = {1\over 2}\sum_f e_f^2\Delta  q_f(x,Q) + O(\alpha_s)\, ,
\label{g1qf}
\eeq
where $\Delta q$ is $q_f^++\bar{q}_f^+-q_f^--\bar{q}_f^-$, the 
difference between the total positive helicity and
negative helicity contributions of quarks and antiquarks of
flavor $f$.  At higher orders, the gluon begins to contribute as well.
The study of polarized parton distributions allows us to
address in parton language
the question of what carries the proton spin.

Polarized distributions
 enjoy the same pattern of factorization
and evolution \cite{Bluemlspin,Robaschik,Stratmann} as
 unpolarized distributions,
may be combined with NLO calculations \cite{Zyla,Ridolfi,Gehrmann,Coriano},
and may be studied 
in the same manner for their small-$x$ behavior  \cite{ErmBart}. 

\begin{figure}
\centerline{\epsfig{file=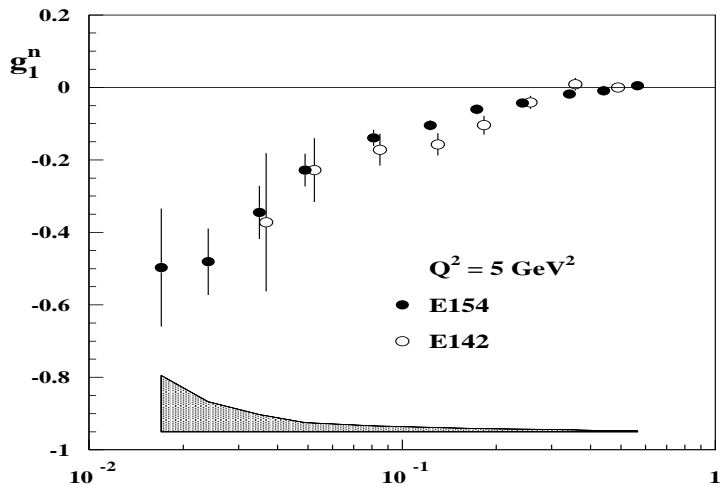,height=3.0in,width=4.5in}}
%\vskip 2.5 true in
%\special{psfile=sfig11.eps hscale=100 vscale=100 hoffset=0 voffset=0}
\caption{Final E154 results for $g_1^n$ {\protect\cite{E154}}.}
\label{E154}
\end{figure}

Within the past two years, efforts have accelerated to
determine polarized parton distributions \cite{Zyla,Ridolfi,wg4}. 
The very recent E154 results for $g_1^n$ (Fig.\ \ref{E154})
will play an enduring
role.  They are also significant
in improving estimates of the neutron contribution to one of the
best-computed predictions of QCD, the Bjorken sum rule,
\begin{equation}
\int_0^1 \left( g_1^p(x)-g_1^n(x)\right)dx = 
{1\over 6}\; \left|{g_A\over g_V}\right|\; \left(1+\sum c_n\alpha_s^n\right)\, ,
\label{BSR}
\end{equation}
with the $c_n$ known out to $c_3$.  
Kinematics, however, restricts the data to a relatively limited range in
$x$.  At the same time, since $g_1$ is decreasing rapidly
as $x\rightarrow 0$, theoretical estimates of
its behavior in this limit \cite{ErmBart} will become
more important as the data becomes more precise in
the accessible region \cite{Stosslein}.  Of course, should HERA itself
become a polarized collider \cite{DeRoeck}, we will
be able to test these predictions. 

Another important current in polarized DIS 
is semiinclusive asymmetries, which have been pioneered
by SMC \cite{Magnon,Kabuss4}, and are now being pushed 
forward by HERMES \cite{Schuler}.  Especially interesting
will be studies of polarized charm production, since it
probes the polarized gluon distribution at leading order.
The caveats mentioned in the
previous section, associated with determining parton distributions
in non-inclusive scattering measurements, apply here as well,
of course.  Certainly, it will be important to complement
measurements of this sort with those from the analysis of
evolution for inclusive structure functions \cite{Robaschik}, and with
the experiments promised at a polarized RHIC.  

\subsection{Extended Polarization Analysis; DVCS}

The parton distribution function $q(x)$ 
may be interpreted as the expectation value 
of the number operator in nucleon state $|p>$,
\beq
q(x)=\int d\ell\; \langle p|\; b^\dagger_q(xp+\ell)\; b_q(xp+\ell)\; |p\rangle\, ,
\eeq
 at
fractional momentum $x$,
integrated over the other momentum components, which we
denote by $\ell$.  The first moment of such a parton distribution,
\beq
p_0\int_0^1 dx\; x\; q(x)
\eeq
is then the contribution of quark $q$ to the total momentum
of the nucleon.  It is, however, not possible to quantify 
contributions to angular momentum in this fashion.  

One of the developments of the preceeding year is
the introduction of a class of ``asymmetric" parton
distribution functions \cite{radyush,Guichon}, in part
to create a partonic language with which to
discuss orbital angular momentum, and in part to 
discuss diffractive vector  boson production, to which 
we will turn shortly.
An asymmetric parton distribution may be thought of 
as a matrix element of the form
\beq
Q(x,x-\delta)= 
\int d\ell\;  \langle (1-\delta)p|\; b^\dagger\left((x-\delta)p+\ell\right)\;
 b(xp+\ell)\; |p\rangle\, .
\label{AsyPDFdef}
\eeq 
Ji \cite{ji} has shown that expectations of this kind are naturally
interpreted in terms of partonic contributions to angular momentum,
and also that the relevant matrix elements are available for study
in what he called deeply virtual Compton scattering (DVCS), 
$p+\gamma^*(Q)\rightarrow p'+\gamma$, the exclusive process in
which an off-shell photon scatters on-shell by exchanging momentum
with a proton.  The appropriate limit for extracting $Q(x,x')$ is one in which the
momentum transfer to the proton is mostly longitudinal, so that
the proton essentially ``slows down" a little bit in the center of 
mass frame.  The observability of this process is under study, and
beyond that questions of evolution and universality may be raised.  
Indeed, as observed above, there is a close relation of DVCS to
the constellation known as diffractive phenomena.  In DVCS 
an on-shell photon is 
produced diffractively.
Let us now turn to the diffractive production of hadrons.

\subsection{Diffraction}

In discussing diffraction, it is worthwhile to keep
in mind the distinction between leading and nonleading
{\it twist}.  Although it has a more exact technical definition,
in DIS the term twist can be used simply
to identify the power behavior of a given
partonic contribution in the
momentum transfer $Q$, leading twist for leading
power in $Q$, and nonleading twist for nonleading power of $Q$.
All the standard unpolarized parton distributions enter at
twist equal to two.  It is possible to 
extend the parton model in perturbative QCD, to include
short-distance reactions that are initiated by {\it more
than one parton}.  

In the standard parton model picture,
we compute cross sections that are the square of
an amplitude for the hard scattering of a single parton.
At nonleading twist, we must consider the possibility
of interference between the amplitude for the hard scattering of a single
parton with  the amplitude for the scattering of two partons
(twist three), or the square of the two-parton scattering amplitude
(twist four).  In general, the twist of such contributions is 
 the number of partons in the amplitude plus
the number in the complex conjugate amplitude.
As we shall see, diffractive processes occur at both leading
and nonleading twist.  

A diffractive DIS process is of the general form
$p+\gamma^*\rightarrow p'+X$, for hadronic final
state $X$ where 
$t=(p-p')^2$ is small.  
Since the proton is present
in both the initial and final states, the production of $X$ involves
no transfer of quantum numbers.
A typical signature
for diffraction is a gap in rapidity between the 
proton and the particles that make up $X$,
but this is not strictly necessary.   When
$t$ is small, convenient kinematic variables are (as usual $W^2=(p+q)^2$)
\beqa
x_{\cal P}&=&{M_X^2+Q^2 \over W^2+Q^2}\nonumber\\
\beta&=&{Q^2 \over 2x_{\cal P}p\cdot q}={Q^2\over M_X^2+Q^2}\nonumber\\
\Rightarrow \beta x_{\cal P}&=&x={Q^2\over 2p\cdot q}\, ,
\label{diffvars}
\eeqa
where $x_{\cal P}$ is the fraction of longitudinal
momentum lost by the proton.  The variable $\beta$ is
an equivalent ``partonic fraction" for the production
of the final state $X$ from the collision of an
object of momentum $x_{\cal P}p$ with the photon
$\gamma^*$.   Since the exchange of momentum to
the photon is free of quantum numbers, it is 
often thought of as carried by the ``pomeron",
a hypothetical object whose exchange is thought to
dominate elastic scattering between
hadrons at fixed momentum transfer and very high energy
(the Regge limit). 

In terms of these variables, it is conventional to define
a four-fold differential cross section, which can be
measured if the momentum of the  final-state proton is actually
observed.
\beq
{d^4\sigma \over dQ^2\; d\beta\; dx_{\cal P}\; dt}
={2\pi\alpha_{\rm EM}^2 \over \beta Q^4}\left(1+(1-y)^2\right) F_2^{D(4)}\, ,
\label{Fdfour}
\eeq
A triply-differential distribution 
is found by integrating over $t$.

Experiments that observe diffraction may be inclusive
or exclusive.  Inclusive diffractive DIS, like inclusive
DIS, may be initiated by a single parton; the soft partons
that carry color into the final state are not observed.
In view of our observations above, inclusive diffraction is
expected to be
leading twist and thus leading power in momentum transfer,
and indeed $F_2^D$ scales in the same manner as the full $F_2$
\cite{Grothe,Dirkmann}.

These considerations suggest a separate factorization
for diffractive processes, analogous to Eq.\ (\ref{otimesfact}),
\beq
F^D=\sum_{\rm partons\ {\it i}} C_i'\otimes f^D_i
\label{diffact}
\eeq
with $f_i^D$ a diffractive parton distribution 
for parton $i$ \cite{Soper,Berera}, defined by the
restriction that the final state include a forward proton.
If this relation holds, we may expect to derive evolution
equations for diffractive as for fully-inclusive DIS.
On the other hand, it is not at all clear that 
diffractive parton distributions are ``universal" in
the same sense as normal distributions \cite{Whitmore}.
After all, a forward proton must survive
very different final-state interactions in, say,
proton-antiproton scattering than in DIS.  
There are growing data on single, and double diffraction
in hadron-hadron scattering with which to compare \cite{Melese,Brandt}.
Data on
the detailed distributions 
of the forward proton in DIS are becoming accessible using
the forward proton and neutron spectrometers at ZEUS
\cite{Cartiglia} and H1 \cite{List,Jansen}.

\begin{figure}[t]
\centerline{\epsfig{file=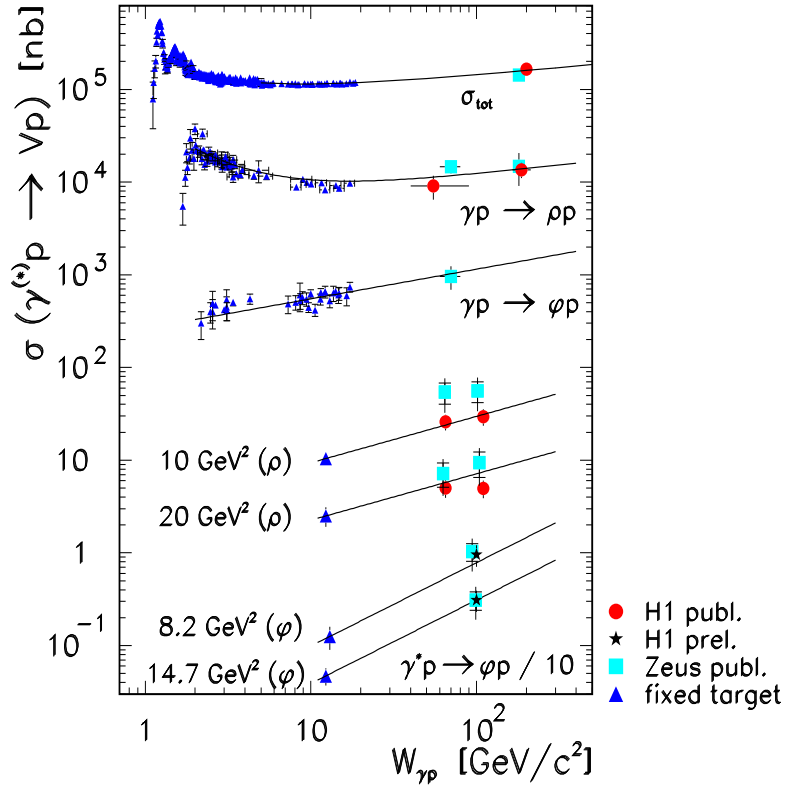,height=4.5in,width=4.5in}}
%\vskip 4.5 true in
%\special{psfile=sfig12.eps hscale=100 vscale=100 hoffset=0 voffset=0}
\caption{$\rho$ and $\phi$ 
production cross section as a function of $W$ for several
values of $Q^2$ {\protect\cite{H1plen}}.}
\label{Rhozeus}
\end{figure}

Some of the most striking data on diffraction are with
exclusive final states, $X=\omega,\ \rho,\ \psi$, etc.
In this case, the perturbative scattering is required
to be color singlet in both amplitude and complex
conjugate.  This requires a minimum of {\it two}
partons for each, so that the exclusive diffraction amplitude begins
at twist four, scaling as $1/M_X^2\sim 1/[(1-x_{\cal P})Q^2]$.
When $M_X$ is large and $t$ is small the amplitude for
this process is described by the same sort of 
asymmetric distribution probed in DVCS above
\cite{Radyushkin2,Freund}. If we
assume that the hard scattering is initiated by the
exchange of two gluons, it is natural to identify this amplitude
with the gluon density, $G(x_{\cal P},M_X)\sim x^{-\lambda(M_X)}$,
where we approximate the steep increase of $G(x,M_X)$ at small 
$x=x_{\cal P}$ and
moderate $M_X$ by a power, increasing with $M_X$
(see Fig.\ \ref{Qfixed}).  Since $W^2\sim 1/x$, 
the cross section, which is proportional to the square of this
amplitude, behaves as
\beq
{d\sigma^D\over dM_X} \sim {1\over M_X^4}W^{4\lambda(M_X)}\, .
\label{diffV}
\eeq
There is now considerable evidence 
in diffractive vector boson production supporting this
general picture \cite{wg2}.  The example of 
Fig.\ \ref{Rhozeus} shows the steepening
of the $W$-dependence of $\rho$- and $\phi$-production for various
values of $Q^2$.

The phenomenology of diffraction has inspired a number of
well-motivated physical models, which should be thought
of as shedding complementary light on this set of processes.  Each was
 extensively discussed at the workshop.  

\begin{figure}[t]
\centerline{\epsfig{file=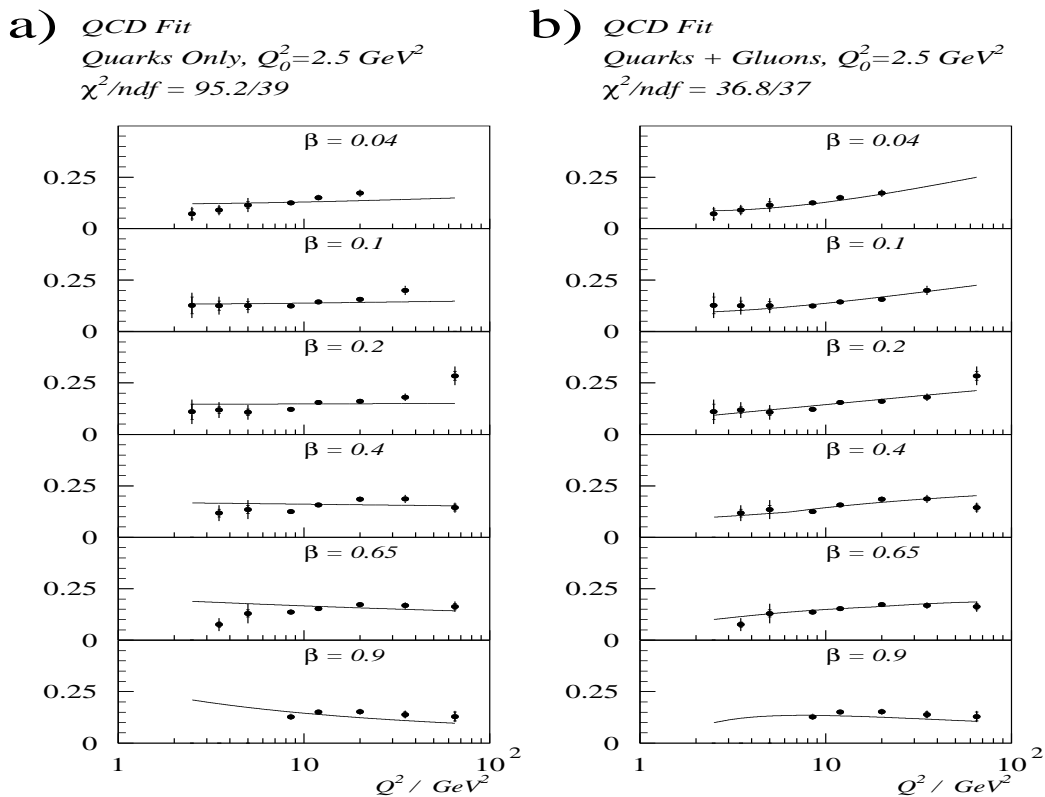,height=4.5in,width=5.5in}}
%\vskip 2.5 true in
%\special{psfile=sfig13.eps hscale=100 vscale=100 hoffset=0 voffset=0}
\caption{QCD fit for the average of $F_2^{(D)}$ (a) with quarks
only, (b) with hard gluon component {\protect\cite{Dirkmann}}.}
\label{avgfd}
\end{figure}

In the model of Ingleman and Schlein \cite{ISch}, hard
diffraction probes the partonic structure of the pomeron,
\beq
f_{a/p}^D=f_{a/{\cal P}}\otimes f_{{\cal P}/p}\, ,
\label{ISotimes}
\eeq
where $f_{a/{\cal P}}(\beta)$ is
the distribution of parton $a$ in the pomeron, while
$f_{{\cal P}/p}$ is the distribution of pomerons
in the proton.  There is as yet no field-theoretic
justification of this picture, which 
differs from Eq.\ (\ref{diffact}) above in assuming
an extra convolution.  Still, it has provided a 
valuable starting-point for the analysis of triply-differential
structure functions \cite{Whitmore}.  These generally
require a very hard gluon distribution in the pomeron,
$f_{g/{\cal P}}(\beta)$, peaked near $\beta\sim 1$.
An example of the evidence for this conclusion is shown
in Fig.\  \ref{avgfd}, in which the average $\int dx F_2^{D(3)}$
is plotted as a function of $Q^2$ \cite{Dirkmann}.  Fits with a very hard
gluon distribution account for the rise with momentum transfer.
In addition, for $x_{\cal P}<1$, fits based on
a pomeron alone seem not to be able to account for
the data, suggesting the need for an additional
coherent object, generically a ``reggeon", $\cal R$, whose
partonic structure is probed in the same fashion,
but whose distribution 
$f_{{\cal R}/p}$ is 
less singular as $x_{\cal P}\rightarrow 1$ \cite{H1plen}.

In the color dipole model \cite{Nikolaev,Wustoff,Bartels,Royon}, 
the photon couples to a quark pair, which interacts 
with the proton by the exchange of two gluons.  
Here the twist, and therefore $Q$-dependence
of the process depends on the polarizations of these
exchanges.  When both gluons carry physical polarizations,
the process is higher-twist, but includes a 
color-singlet component, and can describe exclusive
vector boson production, as above.  When one of the
gluons carries an unphysical polarization, the
process can be leading twist, and is suitable for
describing inclusive diffraction.  The interplay
of polarization and color exchange is particularly well
illustrated in recent work described at the workshop on 
``semiclassical" models, in which, as in the dipole model,
the pair is pictured as interacting with a background
classical color field while passing though the target proton
\cite{McDermott,Hebecker,Buchmuller}.  The simplification
of the coupling of the semiclassical background field
to the color dipole
bears a close relation to  the factorization of soft
gluons from jets in perturbative proofs of factorization theorems
\cite{CSS}.

One of the virtues of the color dipole model is that
it allows the investigation of higher- and leading-twist effects
within a single framework.  Such an analysis \cite{Bartels} suggests 
that as $\beta\rightarrow 1$, higher twist may dominate
for a wide range of $Q$.  If so, the
apparent peak near $\beta=1$
of $f_{g/{\cal P}}(\beta)$, referred to above,
may be the result of a restriction
to a leading twist fit.

\subsection{Rapidity Gaps in Jet Production at the Tevatron}

The interplay between short- and long-distance interactions
is well-illustrated by the rapidity gaps in jet production
observed at the Tevatron.  Unlike the diffractive 
events discussed above, partons from both the 
beams appear to undergo hard scatterings, producing
pairs of jets.  In a detectable fraction of events, however,
there is little or no radiation in the rapidity
range between the jets \cite{Perkins,Bhatti}.  
This is normally attributed to the
exchange of strongly interacting partons in a color singlet
configuration, a ``hard pomeron", which complements the
high-$Q^2$ diffractive vector boson production discussed
above.   These events are more difficult to study, however,
because their large momentum transfers make them rare, and
because they are masked by unconnected ``spectator-spectator"
interactions which reduce their ``survival" in the final state.
We may hope, however, for progress in the theory of
rapidity gap events, as more is learned about the role of color exchange in
hard cross sections \cite{Kidonakis}.

\section{Power Corrections: Openings to Nonperturbative Physics}

As we have noted above, perturbative QCD predictions
at NLO are most accurate when the cross section is
sensitive to only a single large scale.  Jet cross
sections and event shapes, although IR safe, remain
sensitive to hadronization scales even at quite
high energies, as at LEP \cite{DELPHI}.
Event generators have been remarkably successful in
modeling such cross sections, but
they are bound to mask some of the physics in
adjustable parameters.

One of the important developments of the past year
has been a new willingness to step back from event
generators and to confront perturbative computations
directly to the data.  In jet cross sections, the
discrepancy between NLO (or NNLO) theory and experiment
is rather large, of the order of tens of percent, but
shows a characteristic power decay, as $1/Q$ or $1/Q^2$,
depending on the quantity tested.   DIS experiments
are ideal for the study of such effects, since $Q$ 
can be varied in a single experiment.

The impetus for this reanalysis has come from theory
\cite{Beneke5,Dasgupta,Akhoury,Marchesini,Sterman,Sotiropoulos}.
The basic observation is that when we calculate 
an IR safe cross section, we do not eliminate contributions
from soft partons altogether.  Rather, they typically
occur in integrals of the generic form,
\beq
{1\over Q^a}\int_0^Q\; d^{a+2b}k{1\over (k^2)^b}\, .
\label{kint}
\eeq
with $a>0$, where $k$ denotes some set of loop momenta.  Any
integral of this sort is infrared finite, and the region
of soft loop momenta, $k<Q_0$, with $Q_0$ fixed, contributes
a ``power correction", $(Q_0/Q)^a$.   Normally, such corrections
are simply absorbed into the IR safe coefficients of
the strong coupling.   On the other hand, we really do not know
how soft gluons contribute, and we certainly expect that
for $Q_0\sim \Lambda_{\rm QCD}$, nonperturbative scales
will become important.  

\begin{figure}[t]
\centerline{\epsfig{file=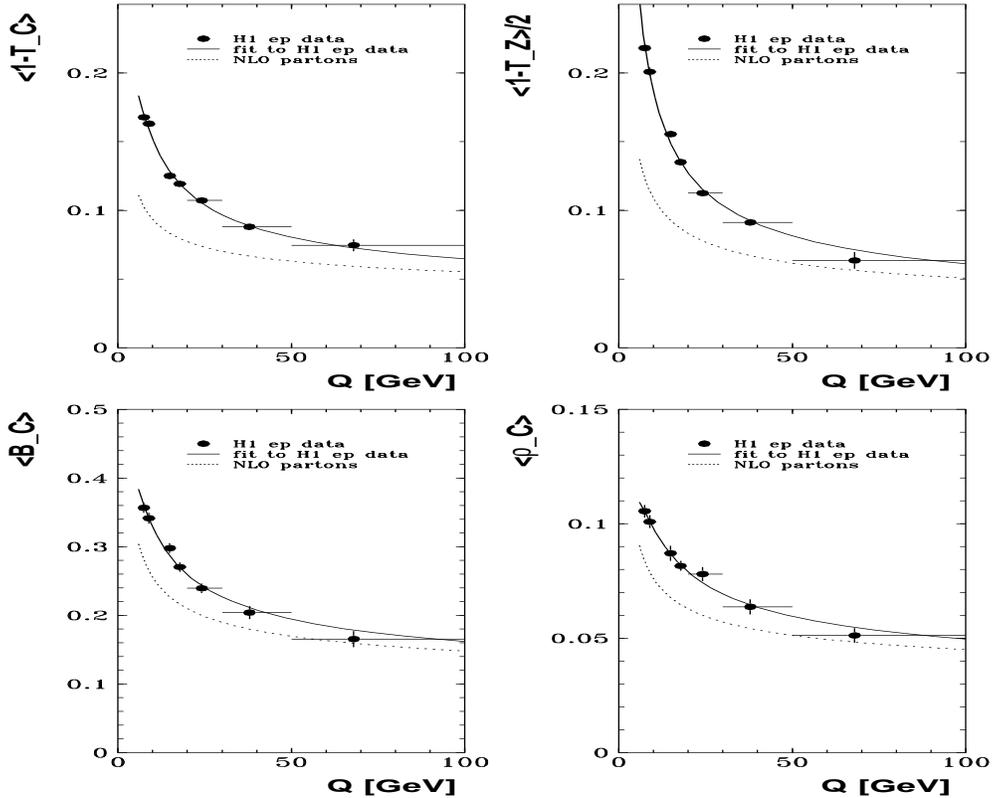,height=4.5in,width=5.5in}}
%\vskip 3.5 true in
%\special{psfile=sfig14.eps hscale=40 vscale=40 hoffset=0 voffset=0}
\caption{Event shapes plotted against $Q$, and compared to
NLO calculations and fits including $1/Q$ corrections {\protect\cite{Rabbertz}}.}
\label{eventshapes}
\end{figure}

For many IRS quantities, it is possible to reorganize
(resum) perturbation theory into a form in which the
running coupling itself signals the introduction
of nonperturbative scales.  Consider, for example, the
thrust $T_C$, defined in DIS as
\beq
T_C={\rm max}_{\hat{n}}\; {\sum_i |p_i\cdot \hat{n}| \over \sum_i |p_i|}\, ,
\eeq
where the sum is over all hadrons 
in the Breit-frame hemisphere of the scattered quark, 
and the maximum is over unit
directions.  $T_C$, which is frame dependent, is maximized to
unity when all the hadrons line
up in a single jet.
At lowest order in perturbation theory, the final
state is a quark pair, and $T_C=1$ exactly.
We thus naturally consider $1-T_C$,
\beq
1-T_C = {1\over Q}\; \sum_i \left ( |p_i|- |p_i \cdot \hat{n}|\right)\, .
\label{1minust}
\eeq
The contribution to $1-T_C$ of a single soft gluon emitted by the
quark is given approximately by 
\beq
\Delta^{(1)}_{1-T_C}=\int {dk_T^2\over k_T^2}\;
C_F{\alpha_s(k_T)\over \pi}\; \int_{k_T}^Q {dk_0\over \sqrt{k_0^2-k_T^2}}
{k_0-\sqrt{k_0^2-k_T^2}\over Q}\, .
\eeq
Here we have let the coupling run with $k_T$, the momentum component
transverse to the quark axis, so that the probability to
emit the extra gluon grows when the gluon becomes either
soft, or collinear to the quark.  The $k_0$ integral
may be carried out explicitly, and leads to an integral of the 
form of Eq.\ (\ref{kint}) above,
\beq
\Delta^{(1)}_{1-T_C}={1\over Q}\; {C_F\over \pi}\; \int_0^Qdk_T\,
\alpha_s(k_T)\, .
\eeq
For $k_T$ large enough, the integrand makes perfect sense, but
when $k_T$ is soft, we are led to modify the perturbative
expression to make it finite, and introduce a new, nonperturbative
parameter with units of mass, which controls the $1/Q$ 
correction to the thrust distribution.  
We have seen at this workshop how expressions derived in this
way can model thrust and other event shape distributions 
at HERA surprisingly well, in terms of only a few new parameters
of this kind.  Examples are shown in Fig. \ref{eventshapes} \cite{Rabbertz}.

Similar analyses have been carried out for ${\rm e}^+{\rm e}^-$ 
annihilation, but the variable hard scattering scale $Q^2$ in
DIS makes it ideal for studying effects of this kind. 
These effects are closely related to nonperturbative contributions
to $k_T$ broadening in $Z$ and Drell-Yan pair production in hadron-hadron
scattering \cite{Montplen}, and undoubtedly to the single-photon spectra
discussed above \cite{Zielinski}.  I believe there is a strong
chance that a unified picture of such effects, cutting across
a variety of processes, will lead to important new insights
in the coming year or two.

Unfortunately, there is no space here to discuss
other theoretical investigations of nonperturbative effects
in high energy processes, including new studies of shadowing, of
the roles of instantons in DIS, and reports of progress
in the lattice computation of parton distributions.  As these
presentations demonstrated, however,
QCD remains a fascinating subject of inquiry at all length scales.

\section{Where are we now?}

What is the place of DIS, the significance of QCD studies?  
Whether or not the HERA high-$Q^2$ events turn out to be a
discovery of historic proportions or
a trick of statistics, they highlight the potential of
deeply inelastic scattering for discovery, and the 
importance of an improved ``standard model" of quantum chromodynamics.
This is certainly a viewpoint that finds a wide resonance
in the physics community.  There is another viewpoint,
 represented at the workshop, although not always stated explicitly.
This is that the study of quantum chromodynamics 
and the investigation of hadronic scattering
are the most challenging problems in quantum field theory that are
currently accessible in the laboratory.  This tradition predates QCD, and
was first posed in the context of a very different theoretical
context.
It was deeply inelastic scattering that brought on a new age --
the age of the parton, then of QCD, which  changed
the set of questions that most theorists wanted to ask. 

It is once again DIS, in a series of advances at HERA, monitored
by the four preceeding DIS conferences, which has led us back to
some of the old questions, which we are finally ready to address again, 
in the new context of quantum field theory.   First the small-$x$
data on structure functions, then the surprising frequency of
diffractive events, the wealth of photoproduction data, all bring
us to the interface of nonperturbative and perturbative
physics, which truly distinguishes QCD as ``the" field theory of the
standard model.  At this interface, our knowledge is very uneven,
progress is halting, and models have an important
role to play.  At the same time, perturbative methods have been
found to be surprisingly, sometimes amazingly, flexible, when the
right questions are asked.
 We must use the QCD we know well to investigate new
physics; but we must also pursue the QCD we do not know well.

\section*{Acknowledgments}

I would like to thank the Organizing Committee, and especially
Jos\'e Repond, for inviting me, and for invaluable help during the
workshop and in the preparation of the manuscript.  I would
also like to thank the Session Convenors of DIS 97 for sharing
many insights, and Vittorio Del Duca, Fred Olness, Anatoly Radyushkin
and Marek Zielinski for very helpful conversations.  This work was
supported in part by the National Science Foundation under
grant PHY 9309888.

\end{document}